# Observation of drastic electronic structure change in one-dimensional moiré crystals

Sihan Zhao[1], Pilkyung Moon[2], Yuhei Miyauchi[3], Kazunari Matsuda[3], Mikito Koshino[4], and Ryo Kitaura[5]*.

[1]Department of Physics, University of California at Berkeley, Berkeley, California 94720, USA

[2]New York University Shanghai, Pudong, Shanghai 200120, China

[3]Institute of Advanced Energy, Kyoto University, Uji, Kyoto 611-0011, Japan

[4]Department of Physics, Osaka University, Toyonaka 560-0043, Japan

[5]Department of Chemistry & Institute for Advanced Research, Nagoya University, Nagoya 464-8602, Japan

*To whom correspondence should be addressed.
 r.kitaura@nagoya-u.jp

**Abstract:** We report the first experimental observation of strong coupling effect in one-dimensional moiré crystals. We study one-dimensional double-wall carbon nanotubes (DWCNTs) in which van der Waals-coupled two single nanotubes form one-dimensional moiré superlattice. We experimentally combine Rayleigh scattering spectroscopy and electron beam diffraction on the same individual DWCNTs to probe the optical transitions of structure-identified DWCNTs in the visible spectral range. Among more than 30 structure-identified DWCNTs examined, we experimentally observed and identified a drastic change of optical transition spectrum in DWCNT with chirality (12,11)@(17,16). The origin of the marked change is attributed to the strong intertube coupling effect in a moiré superlattice formed by two nearly-armchair nanotubes. Our numerical simulation is consistent to these experimental findings.

Engineering the electronic band structures through the formation of moiré superlattice has enabled the discoveries of exotic physics in two-dimensional (2D) van der Waals-coupled heterostructures [1-16]. Such band structure engineering through the formation of moiré superlattice can lead to significant alternation of the electronic properties in the coupled heterostructures. One of the impressive examples is to control the mutual angle between two graphene monolayers to form the twisted bilayer graphene [5-9]. The flat band caused by a moiré superlattice structure gives rise to Mott insulating state and superconducting state at certain twisted "magic" angles, which, on the other hand, do not exist in the pristine monolayers. Moiré superlattice formed between 2D semiconductors also leads to the observation of "moiré excitons" which is a direct consequence of drastic band structure change caused by moiré superlattice [11-16].

While there has been a rapid progress on understanding the moiré physics in 2D heterostructures, experimental studies on moiré superlattice in one-dimensional (1D) systems are still limited, and no clear evidence on strong electronic structure modification has been made and explained in structure-identified 1D moiré superlattices [17-20]. Double-wall carbon nanotubes (DWCNTs), which correspond to a "rolled up" version of twisted bilayer graphene, naturally provide an ideal platform to experimentally probe the moiré physics in 1D. Theoretically, it was assumed that the electronic band structures of realistic DWCNTs are close to those of the individual constituent two single nanotubes, and the effect of the intertube coupling is perturbative [21-24]. Most of previous experimental observations on intertube coupling effect are well-explained within the weakly perturbative regime [25-28]. However, a recent theoretical study predicted that there are special cases where the moiré superlattice potential causes strong coupling between two single nanotubes and can result in completely different band structures [29].

Here we show the experimental evidence of the strong coupling effect in 1D moiré crystals where the moiré superlattice potential significantly alters the electronic band structures in DWCNTs. The lattice structure of each constituent nanotube, consequently the moiré superlattice formed in DWCNT, is determined by electron beam diffraction and the corresponding electronic transitions of the same DWCNT are probed by Rayleigh scattering spectroscopy in the visible spectral range. Though rarely observed in all examined DWCNTs, one specific DWCNTs show drastic changes in Rayleigh scattering spectra, which is assigned to be a direct consequence of the strong coupling between two constituent single nanotubes. The calculated band structure and optical absorption spectra with intertube coupling within the effective continuum model support our experimental interpretation. Our experimental observation of strong coupling effect in 1D DWCNT moiré crystals can open up new opportunities to explore the rich moiré physics in 1D systems.

To investigate the intrinsic moiré physics in DWCNTs, we directly grew suspended DWCNTs with high-quality across an open slit (~ 30 μm in width) by chemical vapor deposition. We studied the individual DWCNTs that are isolated from other nanotubes. The structure (i.e. the chirality) of each constituent single nanotube comprising a DWCNT was determined by the nano-beam electron diffraction with a transmission electron microscope operated at 80 keV (JEOL-2100F). Electronic transitions of the DWCNTs with known chiralities are probed by Rayleigh scattering spectroscopy. In brief, a broadband light from a supercontinuum laser source (1.2 eV–2.75 eV) was focused on the central part of the suspended DWCNT samples and the laser light was polarized along the nanotube axis to probe optical transitions within the same 1D subbands. The light scattered by the nanotube was collected and directed to a CCD camera and a spectrometer. The Rayleigh scattering spectra were obtained by normalizing the measured scattered light with incident laser intensity.

Representative Rayleigh spectra of individual DWCNTs in this study are shown in **Figs. 1a** and **1b**. The chiralities of the two DWCNTs are determined as (16,6)@(22,11) and (23,4)@(22,18) by the nano-beam electron diffraction, respectively. (16,6) and (22,11) in (16,6)@(22,11) correspond to the chirality of inner nanotube and outer nanotube, respectively, and we keep this notation for indexing DWCNTs throughout the manuscript. As clearly seen in the spectra, multiple pronounced optical resonances are observed in both DWCNTs. Each optical resonance in the spectra arises from the dipole-allowed interband transitions within the same 1D subbands from inner and/or outer nanotubes. For example, in (16,6)@(22,11) shown in Fig. 1a, two peaks indicated by the blue arrows are assigned to the $S_{33}$ and $S_{44}$ optical transitions from the outer nanotube (22,11), and that indicated by the red arrow is the $S_{33}$ optical transition from the inner nanotube (16,6). The weak peak at ~ 2.25 eV is the phonon sideband associated with the inner $S_{33}$ transition [30].

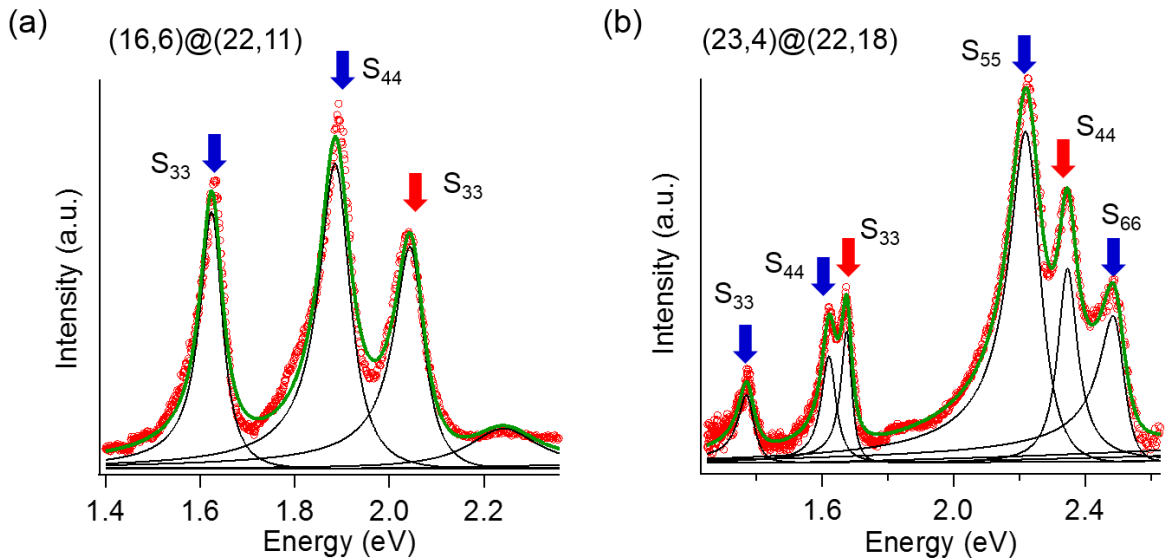

**Figure 1.** Rayleigh scattering spectra of two typical structure-identified DWCNTs with weak-coupling. **(a)** Spectrum for DWCNT (16,6)@(22,11). **(b)** Spectrum for DWCNT (23,4)@(23,18). In both (a) and (b), blue and red arrows mark the optical transitions for inner and outer nanotubes, respectively. The experimental data are presented by red open circles. Each fitted optical transition is presented with black curve and the overall fitted spectrum is shown by solid green line.

To investigate intertube interaction in detail, we precisely determine the resonant transition energies through peak deconvolution with fitting each of the resonances by the form of $I(\omega) \propto \omega^3 |\chi(\omega)|^2$, where $\chi(\omega) \sim A_0 + [(\omega_0 - \omega) - i\gamma/2]^{-1}$. $I(\omega), \chi(\omega), \gamma$, and $A_0$ represent peak intensity, optical susceptibility, full width at half maximum (FWHM) associated with a resonance peaked at $\omega_0$, and non-resonant constant background that accounts for the asymmetric Rayleigh peak shape [31]. The fitted curve of each "Lorentzian-like" optical resonance is shown as the black line and the overall fitted spectrum is presented by the green line. The optical transition energies (i.e. $\omega_0$) for the two DWCNTs are summarized in **Supplemental Material Sec. 1**. Comparing with the transition energies in isolated single nanotubes in air [32], all the optical resonances observed in DWCNTs exhibit noticeable energy redshifts by few tens to one hundred meV primarily due to the dielectric screening caused by the presence of a second nanotube. Note that we observed consistent energy redshifts in most of the examined structure-identified DWCNTs.

While most of DWCNTs investigated are within the weak-coupling regime, where DWCNTs show almost identical electronic transitions to those of each individual constituent nanotube, we experimentally identified two DWCNTs whose electronic transitions are theoretically predicted to be very different from those of constituent single nanotubes due to a non-perturbative intertube coupling between the inner and outer nanotubes[29]. Theoretically, van der Waals coupling between inner and outer nanotubes in DWCNT moiré superlattices is essentially characterized by the relative orientation of chiral vectors (chiralities) of inner and outer nanotubes, C and C'. Significant modification of the band structure takes place under two different conditions, which we call the strong-coupling case and the flat-band case. The former occurs when C and C' are nearly parallel to each other and at the same time the difference of two chiral vectors C' − C is parallel to the armchair direction. Then the moiré superlattice potential makes the resonant coupling between the states of constituent nanotubes, and this leads to drastic energy shift of the subband edges. The latter case occurs under the condition that C and C' are nearly parallel and C' − C is parallel to the zigzag direction. There a long period moiré interference potential turns the original single nanotube bands into a series of nearly flat bands. When C and C' meet either of these two conditions, the electronic structures of DWCNTs become drastically different from the simple sum of constituent single nanotubes [29].

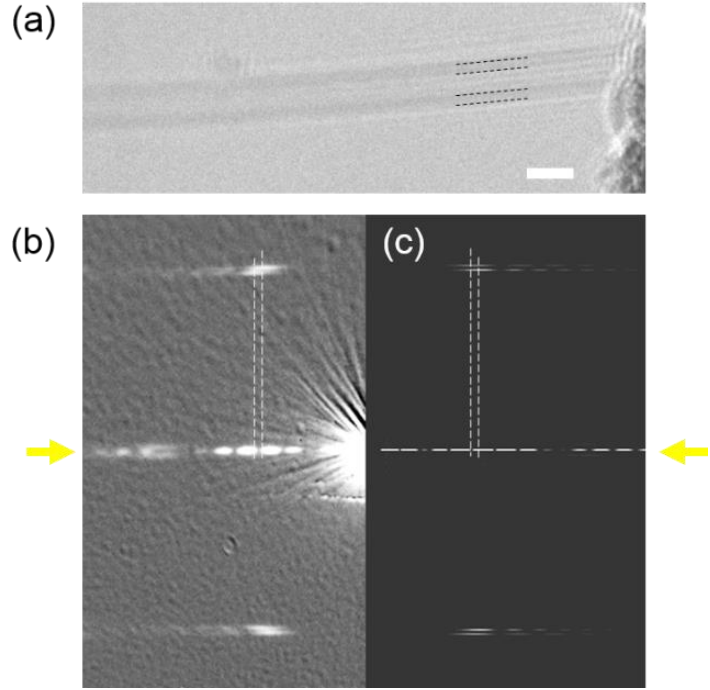

**Figure 2.** Structure characterization of a DWCNT (12,11)@(17,16) showing strong coupling effect. **(a)** TEM image of the DWCNT. Dashed lines indicate the walls of two constituent nanotubes. Scale bar is ~ 2 nm. **(b)** Experimental electron diffraction pattern. **(c)** Simulated electron diffraction pattern of DWCNT with chirality (12,11)@(17,16). Two yellow arrows indicate the position of equatorial line in diffraction pattern. The dashed lines are eye guidance to show one of the key identities between experiment and simulation.

It is easy to see that none of the two weak-coupled DWCNTs shown in Fig. 1 meets the conditions described above. In fact, there is a lot less chance to have a DWCNT meeting the strong coupling criteria. For example, the probability for DWCNTs to satisfy the former criterion is only ~ 0.6% (see **Supplemental Material Sec. 2**). Although we still expect some moderate change of electronic structures in DWCNTs near the criteria, the effects of coupling become weaker as the configuration of C and C' deviates from the criteria. Hereafter we will focus on a strong-coupling case, (12,11)@(17,16) DWCNT, which accurately matches the former criterion. Experimental data for (14,3)@(23,3) DWCNT which approximately matches the latter condition are shown in **Supplemental Material Sec. 3 and Sec. 4**.

**Figure 2a** displays a transmission electron microscopy (TEM) image of a DWCNT in strong coupling regime. Although the TEM image becomes blurred away from the slit edge due to the vibration arising from the suspended structure, it is clear that there are contrasts originating from both inner and outer nanotubes. Judging from the TEM image, this DWCNT comprises an outer nanotube with a diameter ~ 2.2 nm and an inner nanotube with a diameter ~ 1.5 nm with a reasonable intertube distance (~ 0.35 nm). To ambiguously determine the physical structure of each nanotube, electron beam diffraction is employed and an observed diffraction pattern is presented in **Fig. 2b**. Similar to single chiral nanotubes in general, the diffraction pattern of a DWCNT shows sets of mutually twisted hexagonal patterns, which arise from hexagonal lattice of graphitic layers. Prior to any analysis in detail [33], Fig. 2b shows that the two constituent nanotubes are both nearly armchair nanotubes. The equatorial line in the DWCNT diffraction pattern exhibits a "beating-like" oscillation in intensity, which is absent in diffraction patterns in single nanotubes. This unique intensity oscillation originates from the interference of electron waves scattered by two nanotubes in the radial direction. Because of the high sensitivity of the interference to diameter and/or chirality of each nanotube, the oscillation profile along the equatorial line serves as a decisive feature for unambiguous determination of DWCNT chirality. The intensity profile of equatorial line cut in Fig. 2a is further shown in **Supplemental Material Sec. 5**. Based on the analysis procedures reported in literature [34], the chirality for the inner nanotube is determined to be (12,11) and that for the outer nanotube is (17,16). The simulated diffraction pattern of DWCNT (12,11)@(17,16) is shown in **Fig. 2c** for comparison, which shows an excellent agreement with the observed diffraction pattern in experiment. We note that one of the key identities between the Fig. 2b and 2c is indicated by dashed lines where the relative position for features on and outside equatorial line sensitively depend on the detailed chirality of each constituent single nanotubes. We show in **Supplemental Material Sec. 6** the simulated pattern of a DWCNT with chirality (12,11)@(16,15), where chirality of the outer nanotube is slightly different from that of the right chirality DWCNT, (12,11)@(17,16). The discrepancy between experimental result (Fig. 2b) and the simulated pattern of (12,11)@(16,15) is apparent. We systematically examined all the possible candidates with simulation and confirmed that only (12,11)@(17,16) matches the experimental result ( see **Supplemental Material Sec. 7**).

The inner and outer nanotubes in (12,11)@(17,16) are nearly armchair nanotubes, which means that C and C' are nearly parallel to each other (~ 0.4 degree chiral angle difference). In addition, C' − C is (5,5) that is strictly along the armchair direction. (12,11)@(17,16) DWCNT is, therefore, a strong-coupled DWCNT, whose band structure can be strongly modified by 1D Moiré superlattice [29]. The Rayleigh spectrum of the same DWCNT (12,11)@(17,16) is presented in **Fig. 3**. Markedly, we observed total eight well-defined optical resonances over the spectral range 1.35–2.75 eV. On the other hand, within the same photon energy range, pristine (12,11) has two optical transitions $S_{33}$ at ~ 2.23 eV and $S_{44}$ at ~ 2.68 eV ($S_{22}$ at ~ 1.18 eV), and that pristine (17,16) has three optical transitions $S_{33}$ at ~ 1.69 eV, $S_{44}$ at ~ 2.03 eV and $S_{55}$ at ~ 2.66 eV ($S_{22}$ at ~ 0.90 eV) [32]. These referenced energy positions for inner and outer tubes are indicated by the red and blue dashed lines in Fig. 3. Unlike the weak-coupling cases (e.g. Fig. 1), where all optical transitions in DWCNTs correspond to interband transitions of individual constituent nanotubes, (12,11)@(17,16) shows optical resonances that cannot be assigned to those of pristine inner and outer nanotubes. As clearly seen, most of the observed transitions do not match those from each constituent nanotubes. These observations definitely manifest a drastic change of band structure in (12,11)@(17,16), which originates from the strong coupling between the inner and the outer nanotubes. We note that the extra optical transitions cannot be assigned to transitions between different subbands (e.g. $S_{13}$, $S_{24}$ etc.) of each individual constituent nanotubes because of violation of the selection rule with our light polarization direction. We also confirmed that the Rayleigh signals observed in Fig. 3 do not include other nearby nanotubes by carefully examining the structure around the two ends of the DWCNT by TEM.

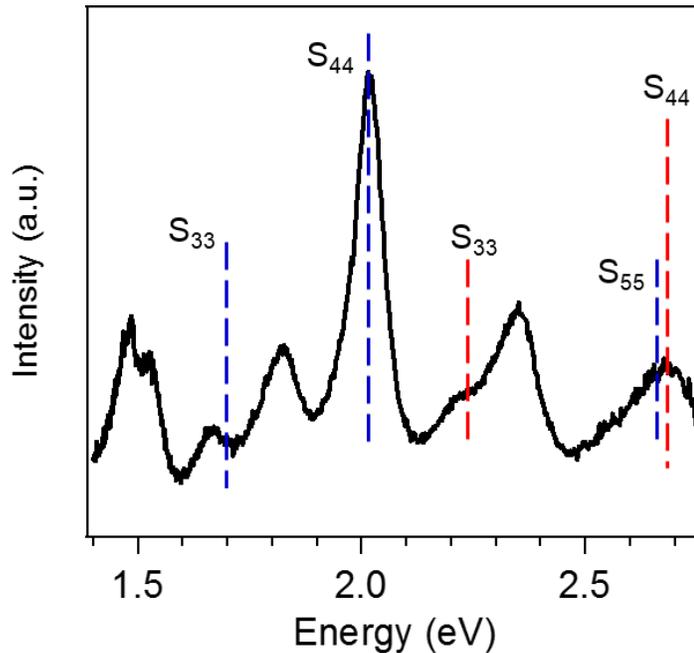

**Figure 3.** Rayleigh spectrum of DWCNT (12,11)@(17,16) showing strong coupling effect. Experimental data are shown in solid black. Dashed red and blue lines indicate the expected optical transition energies for pristine inner and outer nanotubes.

We attribute the observed drastic change of optical transition spectrum to the strong coupling effect arising from 1D moiré superlattice as the theory predicted [29]. By using the effective continuum model in the framework of a tight-binding approximation, we calculated the band structure and optical absorption spectrum for (12,11)@(17,16). The calculated band structure of (12,11)@(17,16) without and with the intertube coupling are shown in **Figs. 4a** and **4b**, respectively. The band structure of pristine inner tube (12,11) and outer tube (17,16) are colored in red and blue in Fig. 4a, respectively. Comparing Figs. 4a and 4b, one can clearly see the high degree of band mixing in the strongly-coupled (12,11)@(17,16) as a consequence of intertube coupling by moiré superlattice potential. **Figure 4c** shows the calculated optical absorption spectrum of the coupled (12,11)@(17,16) (solid black) as well as the absorption spectra of the pristine inner and outer tubes (dashed red and blue). In this energy range, we can see $S_{33}$ and $S_{44}$ of the inner nanotube and $S_{33}$, $S_{44}$, $S_{55}$ and $S_{66}$ of the outer nanotube from the theory (Fig. 4c) whereas, on the other hand, up to outer $S_{55}$ and inner $S_{44}$ are observed in experiment (Fig. 3). We note here that our calculation neglects many-body interactions (electron-electron and electron-hole interactions) and curvature effect, which can lead to significant energy corrections to the one-particle tight-binding results [35]. This explains the position difference of transition energies between experimental results (dashed lines in Fig. 3) and calculated results (dashed red and blue) for pristine single nanotubes. As shown in Fig. 4c, the absorption spectrum of the strongly-coupled DWCNT is very different from sum of the pristine constituent nanotubes. The strong band hybridization caused by the moiré superlattice potential increases the number of optical absorption peaks from five to eight (below $S_{66}$ in Fig. 4c) and makes a large shift of transition energies. Qualitatively, the theoretical calculation captures the main spectral features in Fig. 3, which evidently supports our interpretation on the origin of drastic change of optical transition spectrum observed in DWCNT (12,11)@(17,16). It is the specific moiré superlattice formed between the two nearly armchair (12,11) and (17,16) nanotubes that causes resonant coupling between two nanotubes and results in a drastic change of electronic band structure in this van der Waals-coupled 1D system.

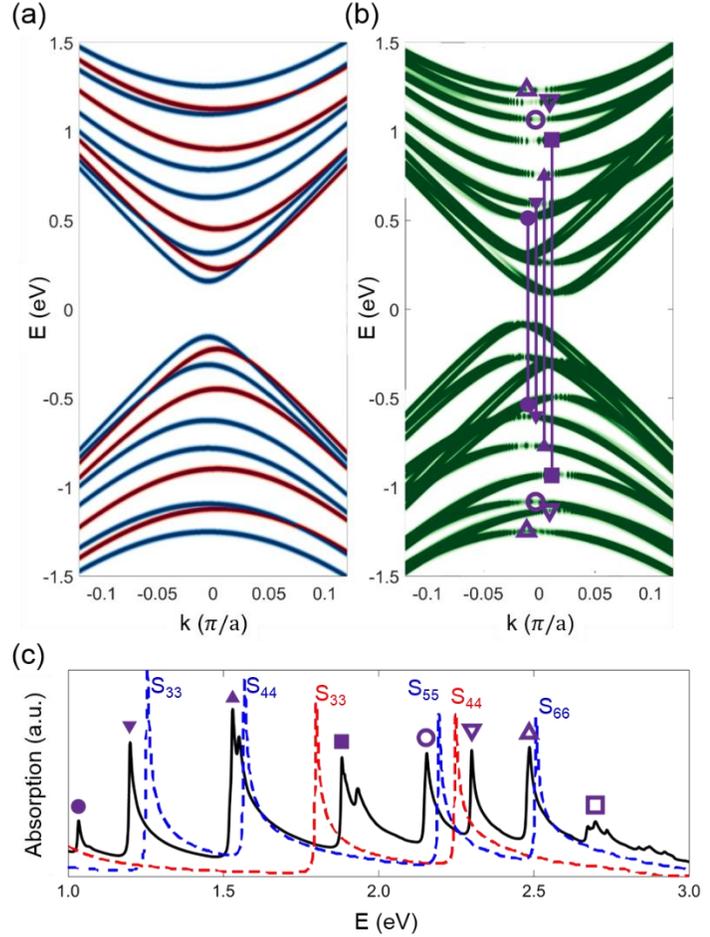

**Figure 4.** Theoretical calculation within tight-binding and effective continuum model. **(a)** Calculated band structure of two pristine nanotubes (12,11) and (17,16) in DWCNT without intertube coupling. **(b)** Calculated band structure of coupled DWCNT (12,11)@(17,16). (c) Calculated optical absorption spectra for pristine inner nanotube (dashed red), pristine outer nanotube (dashed blue) and the coupled DWCNT (solid balck). Optical transitions from pristine inner and outer nanotubes are specified (e.g. $S_{33}$, $S_{44}$ etc.). Different transitions for the coupled DWCNT in (c) are denoted by different makers which originate from interband transitions as indicated with the same makers in the calculated band structure in (b).

In summary, we show the first experimental observation of strong coupling effect in structure-identified 1D moiré crystals. Through combining TEM-based chirality assignment and optical susceptibility measurements with Rayleigh scattering spectroscopy, we have successfully observed optical responses from a strong-coupled DWCNT, (12,11)@(17,16). In contrast to DWCNTs in weak-coupling regime, such as (16,6)@(22,11) and (23,4)@(22,18), the Rayleigh scattering spectrum of (12,11)@(17,16) cannot be understood based on simple sum of individual (12,11) and (17,16) nanotubes. Our numerical simulation shows that the moiré superlatice potential leads to significant alternation of the band structure of (12,11)@(17,16) and the corresponding optical spectra, which is consistent to our experimental observation and interpretation. This is the first definitive experimental observation of moiré superlattice effect in a structure-identified 1D system, which will lead to extended exploration of rich moiré physics in 1D.

**Acknowledgements:** We would like to thank Dr. Seok Jae Yoo in the Department of Physics, UC Berkeley for the helpful discussion. This work was supported by JSPS KAKENHI Grant numbers JP16H06331, JP16H03825, JP16H00963, JP15K13283, JP25107002, and JST CREST Grant Number JPMJCR16F3.

# Supplemental Material

# Observation of drastic electronic structure change in one-dimensional moiré crystals

Sihan Zhao[1], Pilkyung Moon[2], Yuhei Miyauchi[3], Kazunari Matsuda[3], Mikito Koshino[4], and Ryo Kitaura[5]*.

[1]Department of Physics, University of California at Berkeley, Berkeley, California 94720, USA

[2]New York University Shanghai, Pudong, Shanghai 200120, China

[3]Institute of Advanced Energy, Kyoto University, Uji, Kyoto 611-0011, Japan

[4]Department of Physics, Osaka University, Toyonaka 560-0043, Japan

[5]Department of Chemistry & Institute for Advanced Research, Nagoya University, Nagoya 464-8602, Japan

*To whom correspondence should be addressed.

  r.kitaura@nagoya-u.jp

**Supplemental Material Sec. 1:** Summary of transition energies in figure 1 in the main text.

| (16,6)@(22,11) |  | Exp. (eV) | Ref. (eV) [1] | Shift (eV) |
|---|---|---|---|---|
| Inner (16,6) | $S_{33}$ | 2.05 | 2.14 | -0.09 |
| Outer (22,11) | $S_{33}$ | 1.63 | 1.70 | -0.07 |
|  | $S_{44}$ | 1.90 | 1.94 | -0.04 |
| **(23,4)@(22,18)** |  | Exp. (eV) | Ref. (eV) [1] | Shift (eV) |
| Inner (23,4) | $S_{33}$ | 1.68 | 1.76 | -0.08 |
|  | $S_{44}$ | 2.35 | 2.38 | -0.03 |
| Outer (22,18) | $S_{33}$ | 1.37 | 1.42 | -0.05 |
|  | $S_{44}$ | 1.62 | 1.72 | -0.10 |
|  | $S_{55}$ | 2.23 | 2.28 | -0.05 |
|  | $S_{66}$ | 2.50 | 2.61 | -0.11 |

**Supplemental Material Sec. 2:** Note on estimation of strong-coupling probability in DWCNTs. We evaluate the probability strong-coupling case as follows: A DWCNT is denoted by (i,j)@(k,l), where (i,j) represents the outer nanotube and (k,l) stands for the inner nanotube and that i, j, k, l are all integers. To have DWCNTs with reasonable dimeters (< 3 nm for each constituent nanotubes) in general, it requires:

1. i >= j and k >= l
2. Diameter (nm) for outer: $0.246 \times (i^2 + i \times j + j^2)^{0.5} / \pi < 3$
   Diameter (nm) for inner: $0.246 \times (k^2 + k \times l + l^2)^{0.5} / \pi < 3$
3. Intertube distance: $0.33 < \{0.246 \times (i^2 + i \times j + j^2)^{0.5} / \pi - 0.246 \times (k^2 + k \times l + l^2)^{0.5} / \pi\}/2 < 0.38$

With constrictions of 1, 2 and 3, the total number of possible DWCNTs is calculated to be 6549. To have a DWCNT with strong coupling in which inner and outer nanotubes are nearly parallel and the difference of chiral vectors is along armchair direction, it requires to satisfy two extra conditions:

4. Chiral angle difference within ± 1 deg: $-1 < atan((3^{0.5} \times j/(2i+j))) \times 180/\pi - atan((3^{0.5} \times l/(2k+l))) \times 180/\pi < 1$
5. i - k = j - l

With additional 4 and 5, the total number of DWCNTs that satisfy the first criterion to have strong coupling is calculated to be 40. Therefore, the estimated probability to have DWCNT that satisfy the first criterion of strong coupling is 40/6549 ~ 0.6%.

**Supplemental Material Sec. 3:** TEM and electron diffraction for DWCNT (14,3)@(23,3).

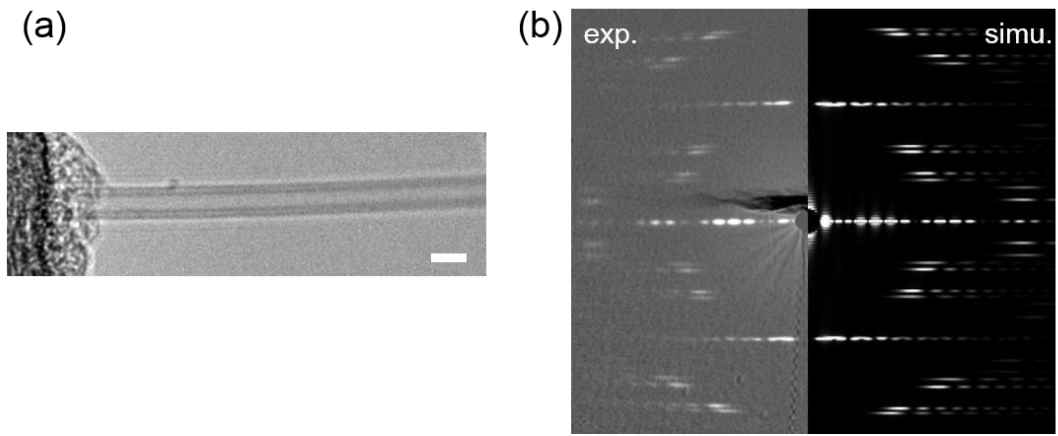

**(a)** TEM image of DWCNT (14,3)@(23,3). Scale bar is ~ 2 nm. **(b)** Experimental diffraction pattern (left) and simulated pattern with chirality (14,3)@(23,3). The experimental result is identical with the simulated pattern.

**Supplemental Material Sec. 4:** Rayleigh spectrum and calculated absorption spectrum for DWCNT (14,3)@(23,3).

The structure of DWCNT (14,3)@(23,3) matches the second coupling condition described in the main text and is predicted to show flat bands as a result of long-period moiré superlattice potential. Experimentally (a), we observed that DWCNT (14,3)@(23,3) shows optical transitions that are close in energy to those of two pristine single nanotubes except for an extra peak (yellow arrow). The three main peaks, $S_{33}$ and $S_{44}$ from the outer tube and $S_{22}$ from the inner tube, shows small energy redshifts of 0.03, 0.08 and 0.03 eV, respectively. On the other hand, our numerical calculation (b) shows that the formation of the flat bands causes peak splitting around the original optical transitions. One possible explanation for this inconsistency is the handedness of CNTs. The strong coupling is expected only when handedness of inner and outer nanotube matches, i.e. right-handed@right-handed or left-handed@left-handed. It is known that electron diffraction cannot distinguish different handedness of nanotubes[2]. If inner and outer nanotubes under this study have opposite handedness, the DWCNT does not show flat band, leading to transitions whose energies are close to those of isolated SWCNTs.

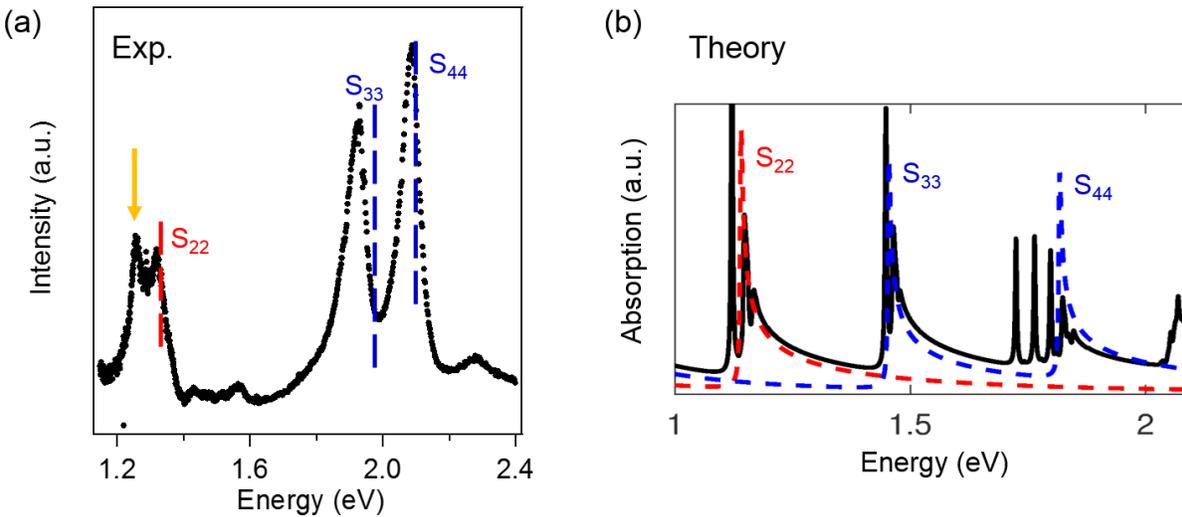

**(a)** A Rayleigh scattering spectrum for DWCNT (14,3)@(23,3). Experimental data is represented by black dots. Energy positions for optical transitions from pristine inner $S_{22}$ and outer $S_{33}$ and $S_{44}$ are respectively indicated by red and blue dashed lines[1]. Additional peak is marked by the yellow arrow. **(b)** Theoretical absorption spectrum for DWCNT (14,3)@(23,3). The calculated spectrum for coupled (14,3)@(23,3) is shown as solid black; those for pristine inner and outer nanotubes are shown by dashed red and blue lines. Note that the energies in (a) and (b) do not precisely match due to the neglect of many-body effect and curvature effect, which can lead to a significant energy correction to single-particle spectrum. Also note that $S_{11}$ for pristine inner nanotube (0.86 eV) and $S_{22}$ for the pristine outer nanotube (0.97 eV) are well below the experimental energy range.

**Supplemental Material Sec. 5:** Intensity profile along equatorial line.

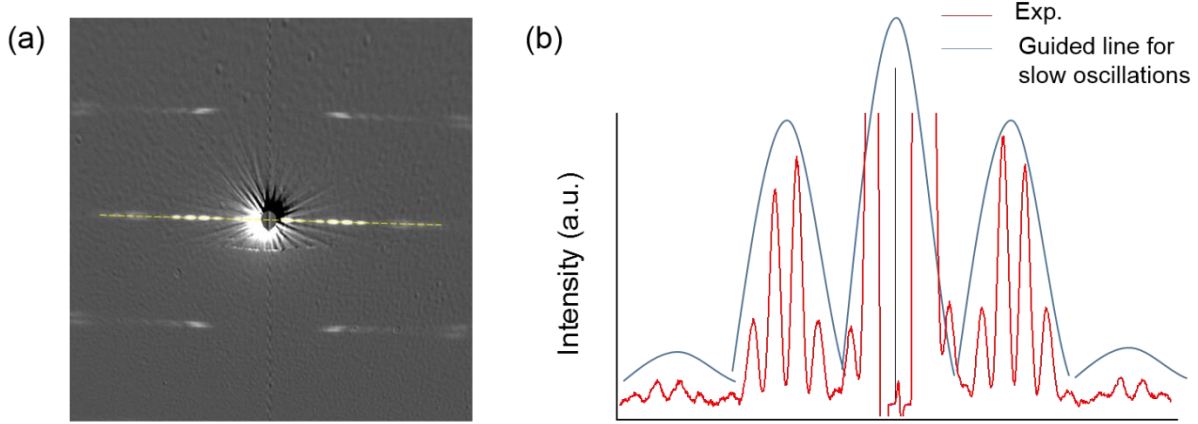

**(a)** Experimental diffraction pattern. Yellow line indicates the line cut along the equatorial line. **(b)** Intensity profile along yellow line in (a). The intensity profile shows "beating-like" fast oscillations along with slow ones (blue curves) which originate from radial interference in inner and outer nanotubes. This "beating-like" oscillations are unique for DWCNTs that are distinctly different from diffraction in single nanotubes.

**Supplemental Material Sec. 6:** Detailed comparison of experimental diffraction pattern with simulated pattern with chirality (12,11)@(16,15).

The simulated pattern in (b) differs from the experimental result in (a) in many aspects. One of the obvious deviations is indicated by the dashed lines where mark the relative position of intensity features on and out of the equatorial line.

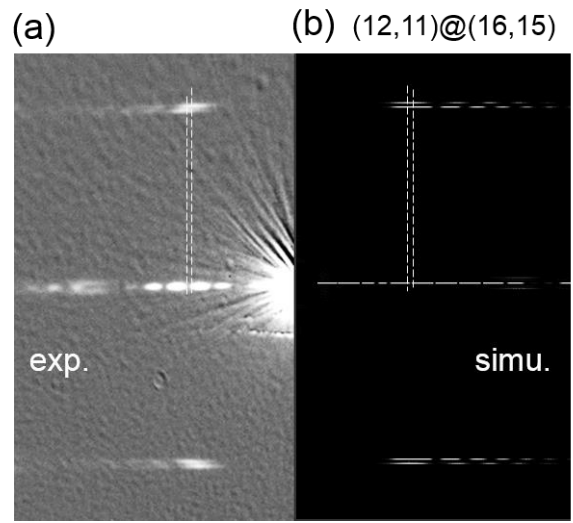

**Supplemental Material Sec. 7:** Systematic comparison of simulated diffraction patterns with experimental result.

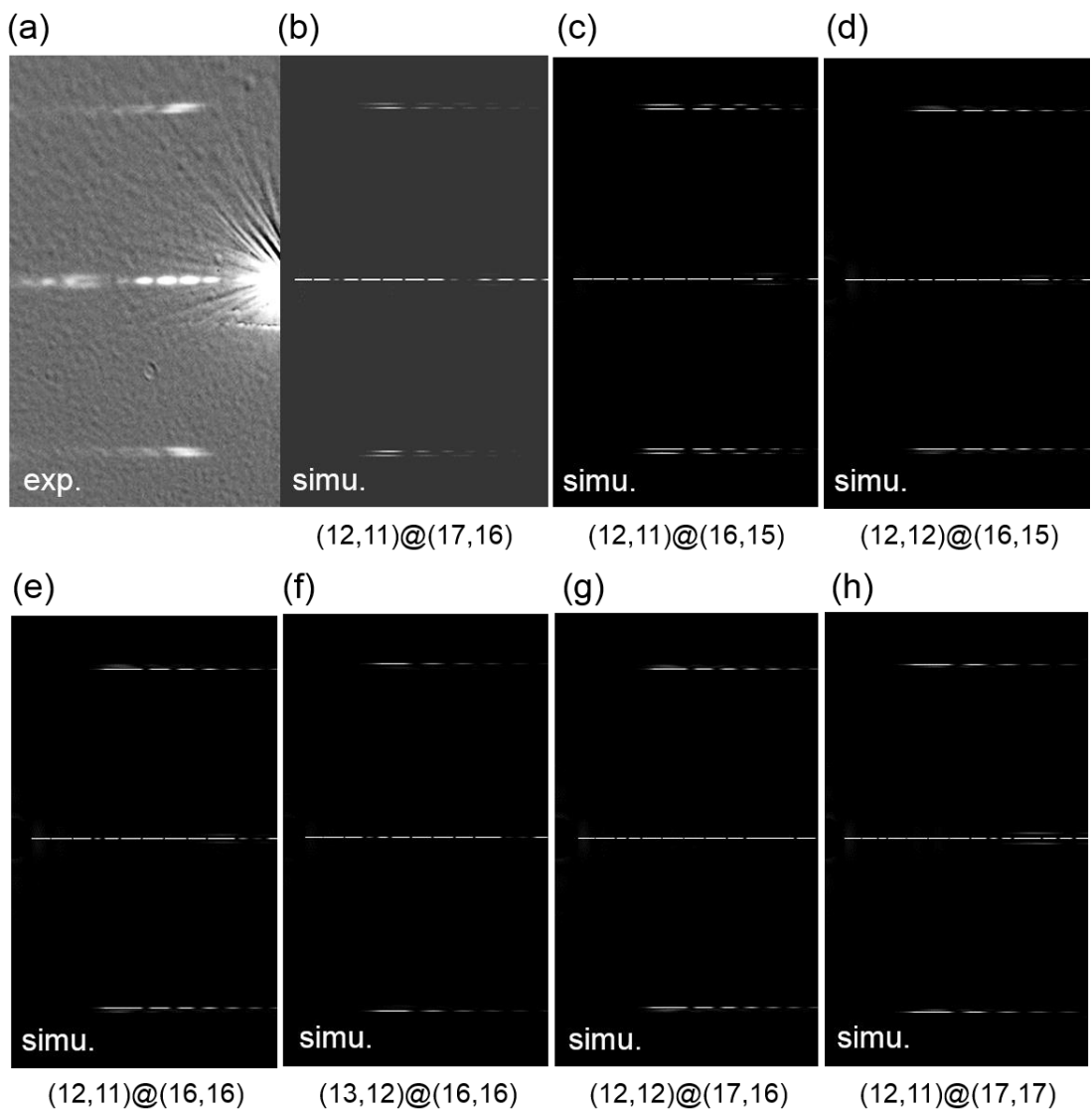

(**a**). Experimental diffraction pattern. (**b**)-(**h**). Simulated DWCNT diffraction patterns with different chiralities. Simulated pattern with (12,11)@(17,16) shown in (b) yields the best matching with the experimental result.